# Electron Correlations in the Low Carrier Density $LaFeAsO_{0.89}F_{0.11}$ Superconductor ($T_c \approx 28$ K)

Athena S. Sefat, Michael A. McGuire, Brian C. Sales, Rongying Jin, Jane Y. Howe, David Mandrus

*Materials Science & Technology Division,*
*Oak Ridge National Laboratory, Oak Ridge, Tennessee 37831, USA*

## Abstract

The crystal structure and numerous normal and superconducting state properties of layered tetragonal (*P*4/*nmm*) LaFeAsO, with F-doping of ≈ 11 %, are reported. Resistivity measurements give an onset transition temperature $T_c = 28.2$ K, and low field magnetic susceptibility data indicate bulk superconductivity. In applied magnetic field, analysis of the resistive transition results in a critical field $H_{c2} \approx 30$ T and a coherence length $\xi_{GL} \approx 35$ Å. An upper limit for the electron carrier concentration of $1 \times 10^{21}$ $cm^{-3}$ is inferred from Hall data just above $T_c$. Strong electron-electron correlations are suggested from temperature-dependent resistivity, Seebeck coefficient, and thermal conductivity data. Anomalies near $T_c$ are observed in both Seebeck coefficient and thermal conductivity data.

## 1. Introduction

Recently, a new oxide family of layered tetragonal superconductors containing the transition metal elements Fe and Ni have been reported: LaFePO ($T_c$ = 4 K, increasing to $T_c$ ~ 7 K with F doping) [1, 2], LaNiPO ($T_c$ = 3 K) [3], and LaFeAsO ($T_c$ = 26 K at 5-11 % F-doping) [4]. These quaternary compounds crystallize with the ZrCuSiAs type structure [5], in *P*4/*nmm* space group (No. 129; Z = 2) [6, 7]. Recent electronic structure calculations [8] describe the structure as sheets of metallic $Fe^{2+}$ in between ionic blocks of $LaOAs^{2-}$. Although there is bonding between Fe and As (d = 2.3-2.4 Å), the states near the Fermi level are dominated by Fe *d*-states lightly mixed with As *p*-states. A view of the structure emphasizing this viewpoint is shown in Fig. 1a. La atoms (with 4*mm* site symmetry) are coordinated by four As and four O atoms, forming distorted square antiprisms. The Fe atoms ($\bar{4}2m$) form square nets perpendicular to the *c*-crystallographic direction.

The focus of this study is the $LaFeAsO_{1-x}F_x$ system with nominal composition of $x \approx 0.11$ [4, 9, 10]. The large reported $T_c$ of approximately 26 K in this compound is very exciting because of the similarities between this layered transition metal compound and the high $T_c$ cuprates. Its discovery may lead to new related superconducting compounds with potentially higher $T_c$'s, but may also provide insight into superconductivity in layered transition metal oxides. The experimental details below are followed by a discussion of the $LaFeAsO_{0.89}F_{0.11}$ crystal structure. The thermodynamic and transport properties of this material will then be presented and discussed. The measurements include temperature dependent specific heat and Hall effect, field- and temperature-dependent magnetic susceptibility, electrical resistivity, Seebeck coefficient and thermal conductivity.

## 2. Experimental details

Polycrystalline samples of $LaFeAsO_{0.89}F_{0.11}$ nominal composition were prepared by a standard solid-state synthesis method similar to that reported by Kamihara *et al.* [4] from elements and binaries with > 4 N. The phase purity and structural identification were made via powder x-ray diffraction, using a Scintag XDS 2000 θ-θ diffractometer (Cu $K_\alpha$ radiation). Data for Rietveld [11] refinement were collected over a 2θ range of 15° to 120° with a step size of 0.02° and a counting time of 10 seconds per step. The data were refined with main phase of $LaFeAsO_{0.89}F_{0.11}$ [4], and small impurity phase of $La_{4.67}(SiO_4)_3O$ [12]. Refined parameters include zero point offset, 10 background polynomial coefficients, two asymmetry parameters, scale factors, preferred orientation, shape and half-width parameters, lattice constants, the two variable atomic positions in $LaFeAsO_{0.89}F_{0.11}$ (*z* coordinates for La and As), and overall isotropic displacement parameters. As expected, the fit was relatively insensitive to the O:F ratio, which was fixed at the nominal value for the final refinement cycles. Atomic positions of the lanthanum silicate oxide phase were not refined, but fixed at the literature values.

Electron probe microanalysis of a polished surface of the polycrystalline pellet was performed on a JEOL JSM-840 scanning electron microscope (SEM) using an accelerating voltage of 10 kV and a current of 20 nA with an EDAX brand energy-dispersive x-ray spectroscopy (EDS) device attached to the scanning electron microscopy (SEM). Transmission electron microscopy (TEM) data were collected on a Hitachi HF-3300TEM/SEM at 300 kV.

DC magnetization was measured as a function of temperature using a Quantum design magnetic property measurement system (MPMS). For a temperature sweep experiment, the sample was cooled to 1.8 K in zero-field (zfc) and data were collected by warming from 2 K to 300 K in an applied field. The sample was then cooled in the applied field (fc), and the measurement repeated from 1.8 K. The magnetic susceptibility results may be presented per mole of $LaFeAsO_{0.89}F_{0.11}$ formula unit ($cm^3$/mol).

Specific heat data, $C_p(T)$, were obtained using a Quantum design physical property measurement system (PPMS) via the relaxation method. The specific heat results may be presented per mole of formula unit (J/K mol) or per mole of atom (J/ K mol atom). Temperature dependent electrical resistivity, thermal conductivity, and Seebeck coefficient measurements were also performed on the PPMS. For dc resistance measurements, electrical contacts were placed on samples in standard 4-probe geometry, using Pt wires and silver epoxy (EPO-TEK H20E). The Hall component was found from the Hall resistivity ($\rho_{xy}$) under magnetic field reversal at a given temperature. For the thermal transport option, gold coated copper leads were used. For thermal transport option, typical $\Delta T/T$ values were about 5% at low temperatures ($T < 40$ K) and about 1% at higher temperatures.

## 3. Results and Discussion

### 3.1 Powder x-ray diffraction and microanalysis

Results of Rietveld refinement of the diffraction data are shown in Fig. 1b. Overall agreement factors are $R_p = 1.36$ and $R_{wp} = 1.86$. The quality of the refinement is satisfactory, as indicated by the value of $\chi^2 = 2.20$. $R_{Bragg}$ for the target phase is low (8.32) indicating a good fit. For the secondary phase of $La_{4.67}(SiO_4)_3O$ the refinement was impeded by its small presence, complexity of the structure, and the overlap between strong peaks of the two phases giving $R_{Bragg} = 24.1$. However, excluding this phase reduces the quality of the overall refinement ($\chi^2 = 3.30$). Quantitative phase analysis was precluded by the poor refinement of the secondary phase. Inspection of a polished surface under optical microscope and SEM (described below) suggests ≈ 5% lanthanum silicate oxide. No evidence of inhomogeneous fluorine distribution is observed in the powder x-ray diffraction peaks.

The refined lattice constants of $LaFeAsO_{0.89}F_{0.11}$ are $a$ = 4.0277(2) Å and $c$ = 8.7125(4) Å, shown in Fig. 1b. These values compare well with that reported for 5 % F-doped sample with $a$ = 4.0320(1) Å and $c$ = 8.7263(3) Å [4], and smaller than the reported lattice constants of LaFeAsO with $a$ = 4.038(1) Å and $c$ = 8.753(6) Å [7]. As expected, the incorporation of F in place of O reduces the size of the unit cell. La and As are located at Wyckoff positions 2*c* with $z$ = 0.1455(3) and 0.6522(5), respectively. Fe is at 2*b* and the shared O/F site is at 2*a*. Nearest neighbor interatomic distances are d(La–O/F) = 2.380(1) Å, d(La–As) = 3.349(3) Å, and d(Fe–As) = 2.411(2) Å. Similar distances are reported for PrFeAsO [7]. The nearest neighbor Fe–Fe distance in $LaFeAsO_{0.89}F_{0.11}$ is 2.8481(1) Å.

Electron probe micro-analysis and SEM are used to complement powder x-ray diffraction data for further investigation of sample purity. Standard-less, semi-quantitative analysis of energy dispersive x-ray spectra confirm the presence of La, Fe, As, O and F with La:Fe:As ≈ 1:1:1. Reliable estimates of the F content are precluded by the near exact overlap of the F-K and Fe-L lines in the microprobe data; however, the presence of F is suggested by the excess intensity of the line near Fe-L, and confirmed independently by x-ray photoelectron spectroscopy measurements. The microprobe analysis reveals the presence of small amounts of two impurity phases: $Fe_2As$ and the lanthanum silicate oxide noted above in the powder diffraction data. No indication of $Fe_2As$ is observed in the powder diffraction data, limiting the concentration of this impurity to less than a few percent. This result is consistent with the magnetic susceptibility data, in that no magnetic anomaly is associated with $Fe_2As$ at ~ 128 K. It is likely that the lanthanum silicate is formed by reaction with SiO vapor present in the silica tubes at the high temperatures of 1250 °C, at which the sample is formed. TEM images and Fast Fourier Transform (FFT) calculated diffraction patterns for two crystallographic orientations of $LaFeAsO_{0.89}F_{0.11}$ are shown in Fig. 1c. The view along the *c*-axis and perpendicular to it are shown on top and bottom, respectively. The lengths of the *a*- and *c*- axes are labeled. The layered nature of the $LaFeAsO_{0.89}F_{0.11}$ structure is clearly visible.

### 3.2 Physical properties

Fig. 2 shows the temperature dependence of the magnetic susceptibility, $\chi$, measured under zfc and fc conditions at 20 Oe. The susceptibility becomes negative below 27.5 K. The shielding fraction is given by the zfc data and the Meissner fraction by the fc curve. Assuming theoretical density of 6.68 g/cm$^3$, a shielding fraction of 74 % and a Meissner fraction of 33 % are found at 2 K. The temperature dependence of susceptibility above $T_c$ at 1 kOe is shown in the inset of Fig. 2. The magnetic susceptibility is similar to that reported by Kamihara et al. [4], with $\chi$ varying from $1.8 \times 10^{-3}$ cm$^3$/mol at room temperature to $4.1 \times 10^{-3}$ cm$^3$/mol at 30 K, with M vs H curves linear in this temperature range. This value is approximately 50 times large in magnitude compared to the expected bare susceptibility of $\chi_o = 8.51 \times 10^{-5}$ cm$^3$/mol, recently found from density functional theory and local spin density approximation (L(S)DA) [8]. Several different fluorine-doped LaFeAsO samples were synthesized using different combinations of elements and binary phases, but the high temperature susceptibility data from all of these samples were of similar magnitude to that shown in Fig. 2, inset. This suggests that the large value of the susceptibility may be intrinsic to the superconducting phase, and not due to an impurity phase or inhomogeneous fluorine doping (see Section 3.1). This indicates that the superconductor may be close to magnetic ordering. Within the LDA, LaFeAsO is also reported to be on the borderline of a ferromagnetic instability [8].

Fig. 3a shows the temperature dependence of the electrical resistivity ($\rho$) in zero field. At room temperature $\rho_{300\ K} = 2.7$ m$\Omega$ cm, comparable to the reported values of 3.5 m$\Omega$ cm [4] and 2.2 m$\Omega$ cm [5]. The inset is the enlarged view below 50 K with the onset transition temperature $T_c^{onset} = 28.2$ K and a transition width $\Delta T_c = T_c (90\%) - T_c (10\%) = 4.5$ K. In analyzing the resistivity data above $T_c$, we find that $\rho$ exhibits a quadratic temperature dependence, $\rho = \rho_o + AT^2$, over a wide temperature range. Plotted in Fig. 3b is $\rho(T^2)$ for H = 0 (open triangles) and H = 8 T (filled circles), and linear fits between $T_c$ and 200 K. The *A* values are $3.5 \times 10^{-5}$ m$\Omega$ cm K$^{-2}$ and $3.4 \times 10^{-5}$ m$\Omega$ cm K$^{-2}$ for H = 0 and 8 T, respectively, and $\rho_o \approx 0.11$ m$\Omega$ cm. The $T^2$ behavior of $\rho$ below 200 K indicates the importance of the Umklapp process of the electron-electron scattering and is consistent with the formation of a Fermi-liquid state. The *A* values extracted here are comparable with semi-heavy-Fermion compounds such as $CePd_3$ and $UIn_3$ [13].

Fig. 4a illustrates that the resistive transition for $LaFeAsO_{0.89}F_{0.11}$ shifts to lower temperatures by applying a magnetic field. The transition width becomes wider with increasing H, a characteristic of type-II superconductivity. Here we define a transition temperature $T_c(H)$ which satisfies the condition that $\rho(T_c, H)$ equals a fixed percentage of the normal-state value ($\rho_N$) for each field H. The $T_c(H)$ values for $\rho$ = 10, 50, and 90 % are shown in Fig. 4b, represented by the upper critical field $H_{c2}(T)$. In all cases we find that $H_{c2}(T)$ has a linear dependence with no sign of saturation. The slope $(dH_{c2}/dT)|_{T=Tc}$ = -0.87 T/K for $\rho_N$ = 10%, -1.41 T/K for $\rho_N$ = 50 %, and -1.59 T/K for $\rho_N$ = 90 %. In the conventional Bardeen-Cooper-Schrieffer (BCS) picture, $H_{c2}$ is linear in T near $T_{c0}$ and saturates in the 0 K limit, however deviations may be caused in the presence of impurity scattering [14]. In the latter case, Werthamer-Helfand-Hohenberg (WHH) equation is $H_{c2}(0) = -0.693\ T_c(dH_{c2}/dT)|_{T=Tc}$. The dashed lines in Fig. 4b are the results of fitting $H_{c2}(T)$ to the WHH formula, yielding $H_{c2}^{WHH}(0)$ = 14.3 T for $\rho_N$ = 10%, 25.0 T for $\rho_N$ = 50 %, and 31.2 T for $\rho_N$ = 90 %. It should be noted that at lower temperatures $H_{c2}(T)$ no longer follows the WHH expression, particularly for $\rho_N$ = 10%, which suggests that $H_{c2}(0)$ is probably larger than $H_{c2}^{WHH}(0)$. In a recent publication, higher upper critical field of ~ 54 T was reported [5]. Nevertheless, assuming $H_{c2}(0) = H_{c2}^{WHH}(0)$, Ginzburg-Landau formula is $\xi_{GL} = (\Phi_0/2\pi H_{c2})^{1/2}$, where $\Phi_0 = 2.07 \times 10^{-7}$ Oe cm$^2$. This yields zero temperature coherence length $\xi_{GL}(0) \approx 48$ Å for $H_{c2}$(10% $\rho_N$), 36 Å for $H_{c2}$(50% $\rho_N$), and 33 Å for $H_{c2}$(90% $\rho_N$). These values are slightly larger than that reported (25 Å) [5], and comparable to those for high-temperature superconducting cuprates with similar transition temperatures [15-17].

The temperature dependent specific heat for $LaFeAsO_{0.89}F_{0.11}$ is shown in Fig. 5a, with enlarged low temperature region in the inset showing the lack of an anomaly near $T_c$. The large value of the lattice specific heat near $T_c$ would make the superconducting contribution at $T_c$ fairly small (of order 1-3%), but the lack of any anomaly above the precision of our data is somewhat surprising. We measured several different polycrystalline samples that showed relatively large Meissner fractions, but were unable to detect a specific heat anomaly at $T_c$. The C/T versus $T^2$ plot (Fig. 5b) follows a straight line for the region of ~4 to 8 K, which allows the estimation of electronic $\gamma$ and lattice $\beta$ values. The contributions are $\gamma$ = 1.0(3) mJ/K$^2$mol atom [4.1(1) mJ/K$^2$mol], and $\beta$ = 0.139(1) mJ/K$^4$mol atom. The $\gamma$-value may not be intrinsic to the superconducting phase. The value of the Debye temperature ($\theta_D$) can be calculated from $\beta = 12\pi^4 R/5\theta_D^3$ at low temperatures and resulting in $\theta_D \approx 240$ K. From high temperature data, $\theta_D \approx 325$ K; this value is comparable to that recently derived from specific heat results ($\theta_D$ = 315.7 K) [10].

Fig. 6 shows the results of temperature dependent Hall data. The Hall coefficient (Fig. 6, inset) is negative and decreases gradually with decreasing temperature down to ~ 130 K below which it is approximately constant. If a single band is assumed, the inferred carrier concentration is $\approx 1.7 \times 10^{21}$ electrons/cm$^3$ at room temperature and $\approx 1 \times 10^{21}$ electrons/cm$^3$ just above $T_c$. Our Hall data are similar to those reported recently [9]. The electronic structure calculations [8] also predict a similar carrier concentration, as well as the presence of both high velocity electron bands and heavy hole bands near the Fermi energy, with the electron bands dominating in-plane transport. Because of the possibility of some electrical conduction by holes, the low temperature carrier concentration of $1 \times 10^{21}$ electrons/cm$^3$ should be regarded as an upper bound for the electron carrier concentration just above $T_c$.

The Seebeck coefficient of $LaFeAsO_{0.89}F_{0.11}$, S, is shown in Fig. 7. S is negative and large, varying from -42 μV/K at 300 K to a maximum negative value of ~ –95 μV/K at ~ 130 K, then decreasing toward 0. The maximum in the Seebeck coefficient and the slight temperature dependence of the Hall data is probably due to the competition between dominant electron-like bands and the expected proximity of hole-like bands near the Fermi energy.

The low temperature Seebeck and Hall data (Fig. 6 & 7) can be used to obtain a qualitative estimate of the effective mass of the carriers, $m^*$, and the Sommerfeld coefficient, γ. If we assume that the electron carrier concentration below ~ 130 K is close to the value inferred from the Hall data (1 x $10^{21}$ electrons/$cm^3$), the free electron model predicts a Fermi energy (temperature) of 0.4 eV (4800 K) [19]. The Seebeck coefficient at 40 K is –60 μV/K and in the temperature regime, roughly extrapolates linearly to 0 at T = 0 (Fig. 7). In many normal low carrier concentration metals [18], the Seebeck coefficient is given by the diffusion contribution of the Mott expression or S = $\pi^2 k_B T(2eT_F)^{-1}$, where $T_F$ is the Fermi temperature. If we use a value for of 4800 K for $T_F$ in the expression for S, this results in a Seebeck value of ≈ -3.5 μV/K at 40 K, which is much smaller than the measured value of –60 μV/K. These results imply a value for $m^* \approx 17$ and a corresponding value of $\gamma \approx 11$ mJ/$K^2$ mol [19], which is about twice the calculated density of states at $E_f$ ($\gamma_o$ = 6.5 mJ/$K^2$ mol) [8]. Although this is clearly a crude estimate, it suggests that electronic correlations are important in $LaFeAsO_{0.89}F_{0.11}$, as were suggested from $\rho(T^2)$ analyses above. We also note that the value of γ = 4.1(1) mJ/$K^2$mol extracted from the specific heat data may reflect that a portion of the $LaFeAsO_{0.89}F_{0.11}$ phase remains in the normal state.

For $LaFeAsO_{0.89}F_{0.11}$, the temperature dependent thermal conductivity is displayed in Fig. 8. An anomaly at ≈ 28 K is clearly evident. Because of the low carrier concentration and high resistivity, most of the heat in this compound is carried by phonons. Above $T_c$, approximately 85 % of the heat is carried by phonons, as estimated using the Wiedemann-Franz relationship. Just below $T_c$, the electrons that condense into the superconducting state should carry no heat because of the opening of the superconducting gap. Since electrons that carry heat are removed below $T_c$, one would expect that the thermal conductivity should decrease below $T_c$, as occurs in most superconductors [20-22]. The opposite occurs in $LaFeAsO_{0.89}F_{0.11}$ below $T_c$, as the thermal conductivity is *higher* than the expected normal state values. This is clear from a comparison of the 0 field data with thermal conductivity data taken at 8 T that suppresses $T_c$ by about 5 K (Fig. 8). Similar behavior is observed in the cuprate superconductors [23, 24]. One explanation for the higher value of κ for $LaFeAsO_{0.89}F_{0.11}$ just below $T_c$, consistent with the evidence for strong electron-electron scattering, is an increase in the mobility of the fraction of electrons still in the normal state.

## 4. Conclusions

The crystal structure, magnetic susceptibility, specific heat, resistivity, Hall effect, Seebeck coefficient and thermal conductivity of the layered $LaFeAsO_{0.89}F_{0.11}$ superconductor were investigated in this report. The magnetic susceptibility clearly indicates bulk superconductivity with Meissner fraction of ≈ 30 % and screening fraction of ≈ 70 %. Analysis of the resistivity, Seebeck coefficient, thermal conductivity and high temperature magnetic susceptibility data suggest that electron-electron correlations are large in this compound. The relationship between correlations and superconductivity will require further experimental investigations using single crystals as well as substantial insight from theoretical calculations.

## Acknowledgement

We would like to thank D. J. Singh for helpful discussions, H. M. Meyer for XPS analyses, and G. M. Veith for assistance with SEM and x-ray diffraction data collections. Research sponsored by the Division of Materials Science and Engineering, Office of Basic Energy Sciences. Oak Ridge National Laboratory is managed by UT-Battelle, LLC, for the U.S. Department of Energy under Contract No. DE-AC05-00OR22725.

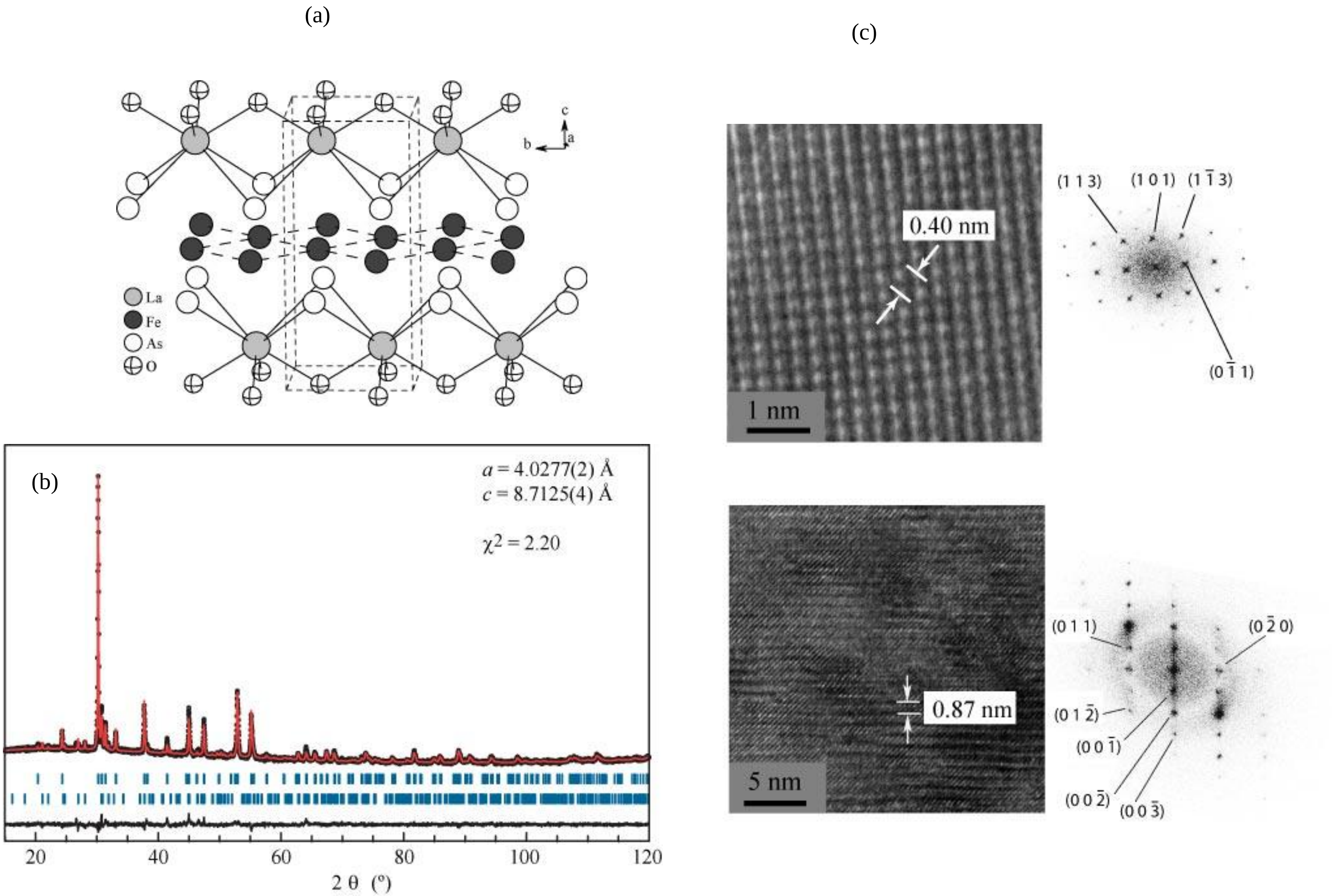

Figure 1: (a) Crystal structure of $LaFeAsO_{0.89}F_{0.11}$, with unit cell shown in dotted lines. The structure is made from layers of face-sharing distorted square antiprism of $LaAs_4O_4$ and sheets of Fe, perpendicular to *c*-direction. (b) The refined x-ray powder diffraction of nominal composition $LaFeAsO_{0.89}F_{0.11}$. The upper set of Bragg ticks correspond to reflections from this structure; the lower ticks locate reflections from ≈ 5% impurity phase of $La_{4.67}(SiO_4)_3O$. (c) TEM images and FFT calculated diffraction patterns for two crystallographic orientations of $LaFeAsO_{0.89}F_{0.11}$. The view along the *c*-axis and perpendicular to it are shown in the top and bottom panels, respectively. The lengths of *a*- and *c*- axes are labeled.

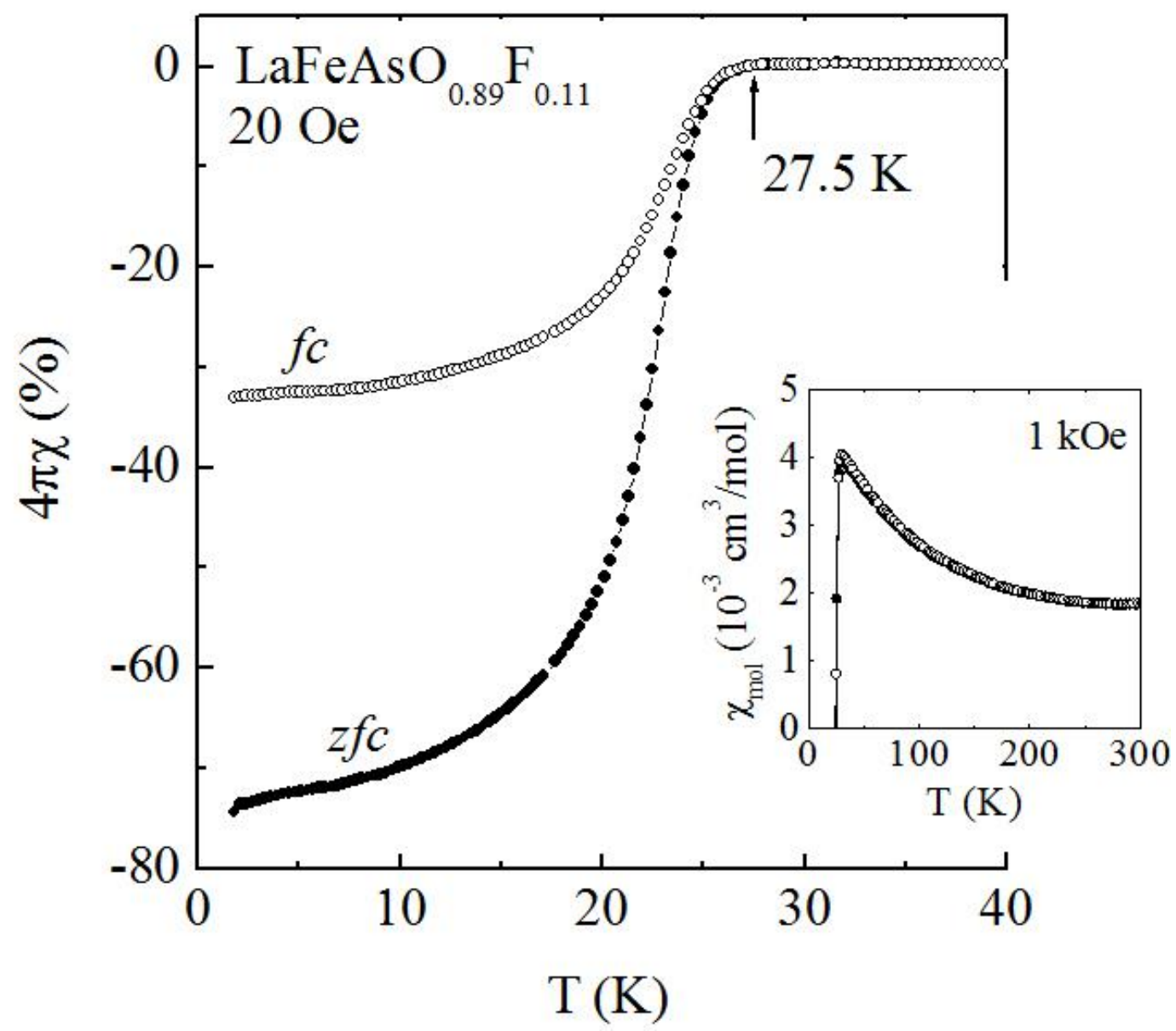

Figure 2: Temperature dependence of the zfc (filled circles) and fc (open circles) magnetic susceptibility for $LaFeAsO_{0.89}F_{0.11}$ in 20 Oe and 1 kOe (inset).

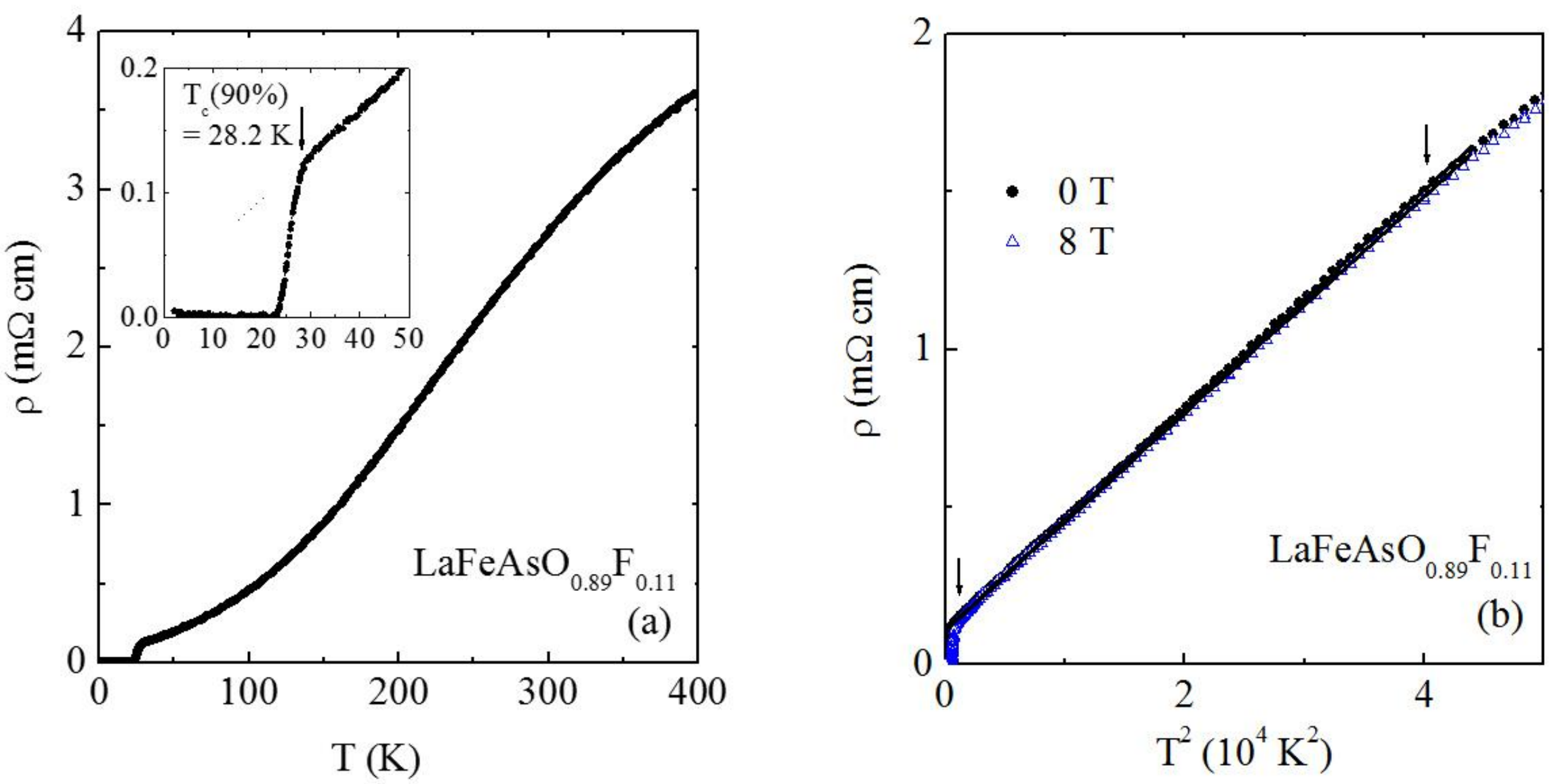

Figure 3: Temperature dependence of electrical resistivity for $LaFeAsO_{0.89}F_{0.11}$, shown as (a) ρ vs T and (b) ρ vs $T^2$ between 0 and ~ 225 K. Inset of (a) is the enlarged low temperature data, with the indicated superconducting transition temperature. In (b) the red solid lines represent the linear fit between 35 and 200 K (shown by arrows) for data in 0 and 8 T.

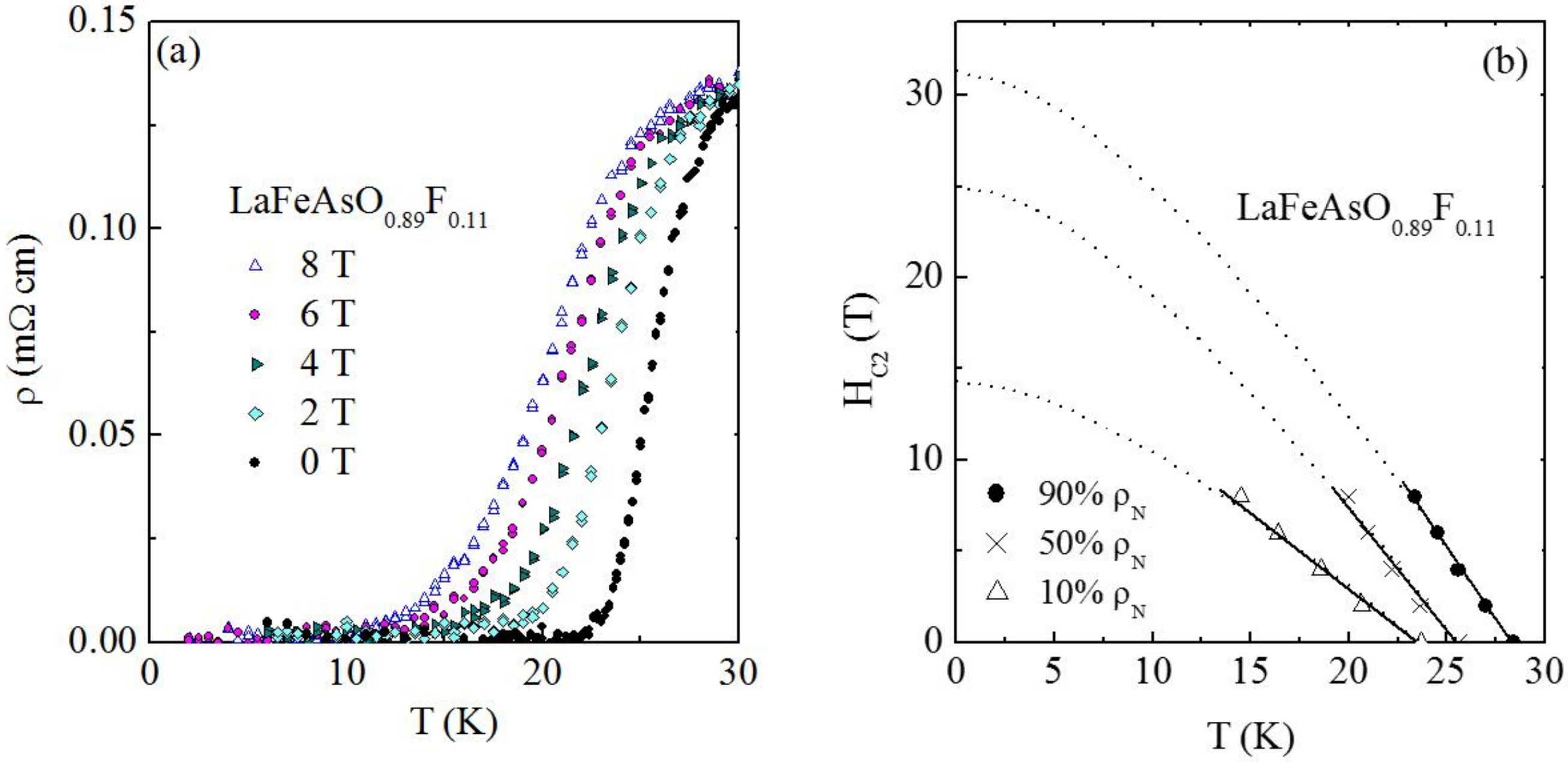

Figure 4: For $LaFeAsO_{0.89}F_{0.11}$, the temperature dependence of (a) resistivity at various applied fields and (b) the upper critical field $H_{c2}$ found from 90%, 50%, and 10% of the normal-state value, $\rho_N$. In (b), the dotted lines represent the WHH approach and the solid lines are the linear fits to experimental $H_{c2}(T)$.

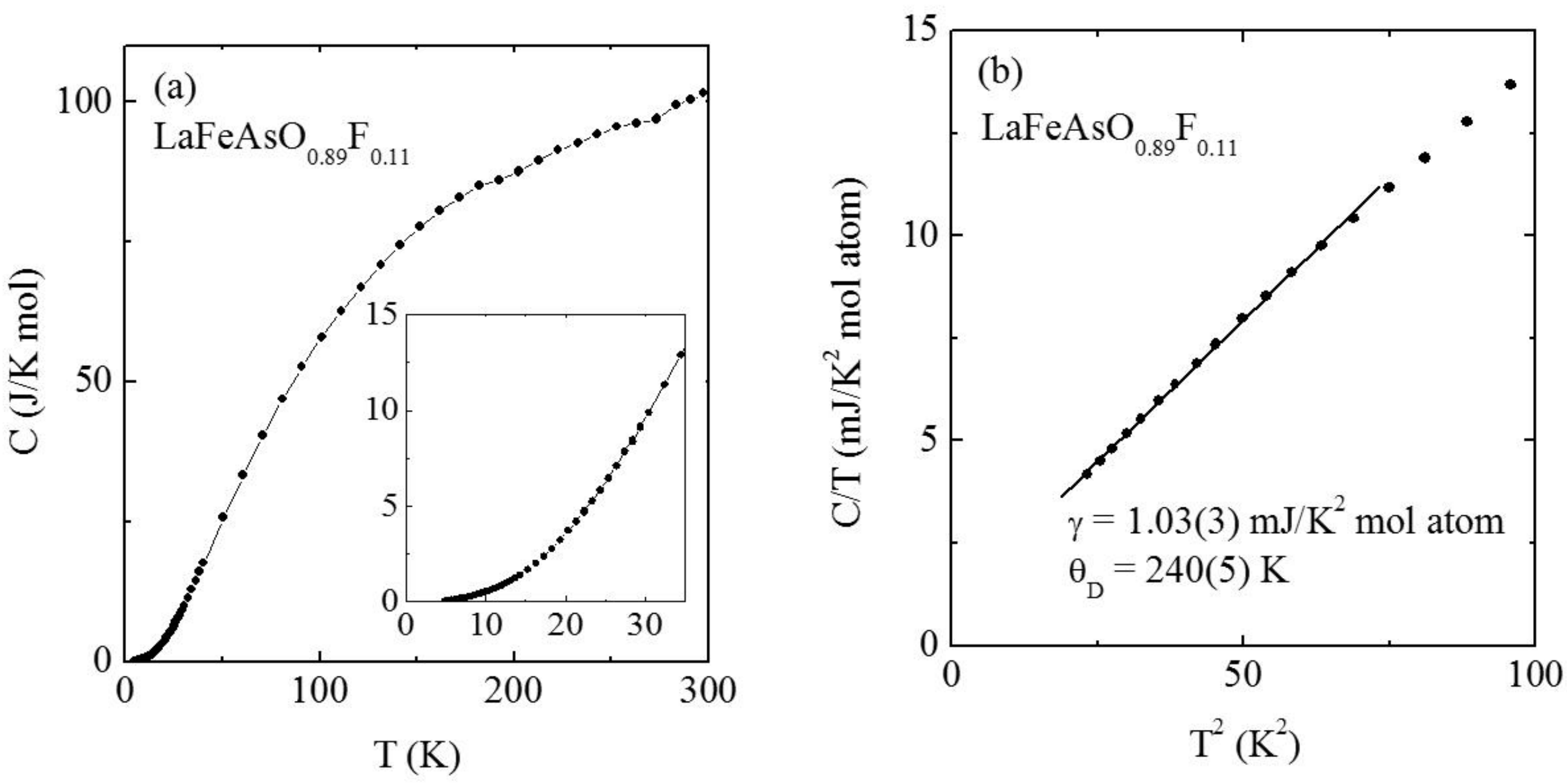

Figure 5: Temperature dependence of specific heat for $LaFeAsO_{0.89}F_{0.11}$ (a) C versus T and (b) C/T vs $T^2$ and linear fit between ~ 4 and 8 K. The inset of (a) shows the lack of anomaly near $T_c$.

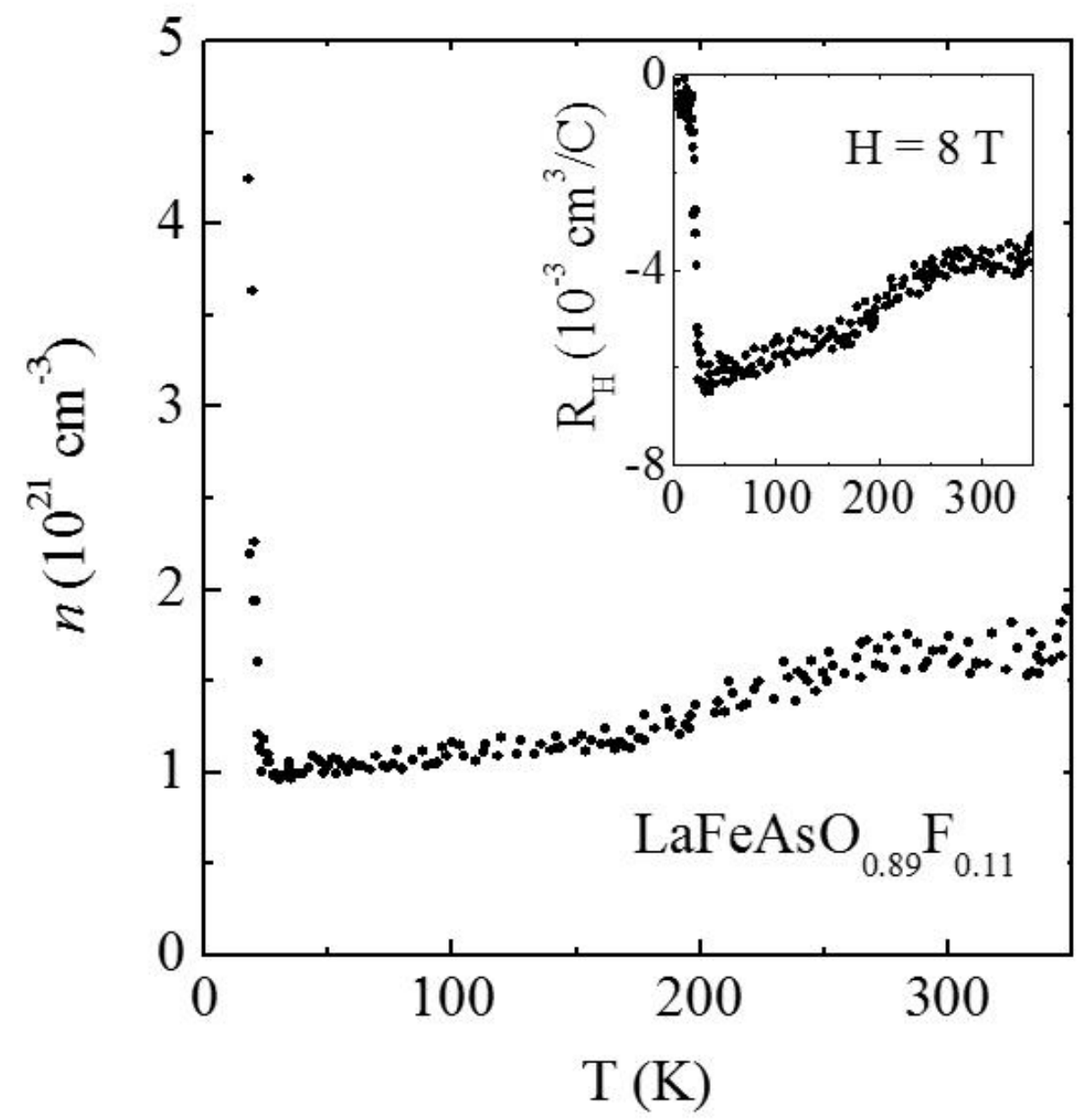

Figure 6: The inferred variation of the carrier density *n* with temperature for $LaFeAsO_{0.89}F_{0.11}$. The inset is the temperature dependence of the Hall coefficient $R_H$ obtained in a magnetic field of 8 T.

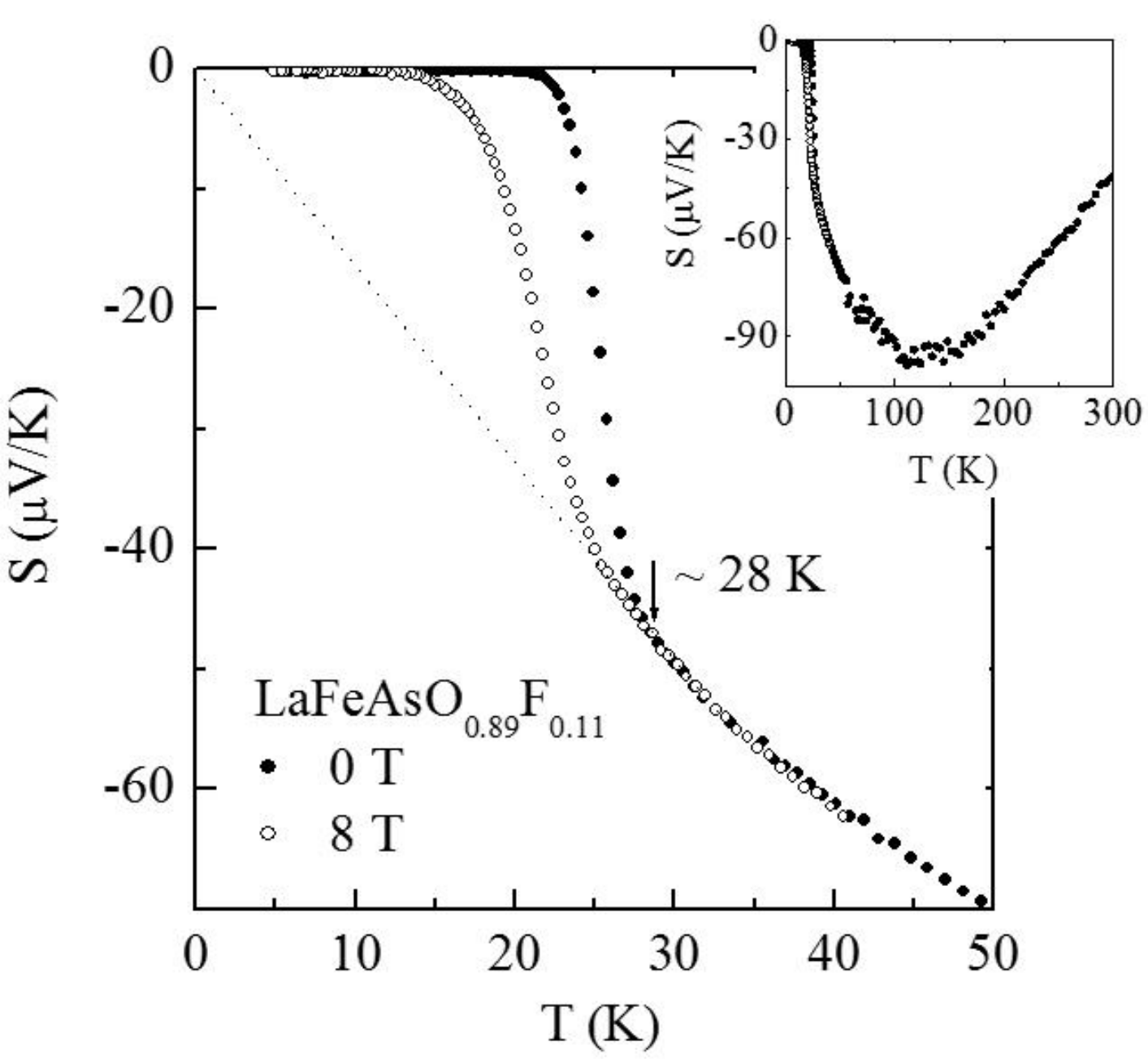

Figure 7: Temperature dependence of Seebeck coefficient for $LaFeAsO_{0.89}F_{0.11}$ in applied fields of 0 and 8 T. The inset shows the temperature dependence up to room temperature. The dotted line is extrapolation from ~ 35 K down to 0.

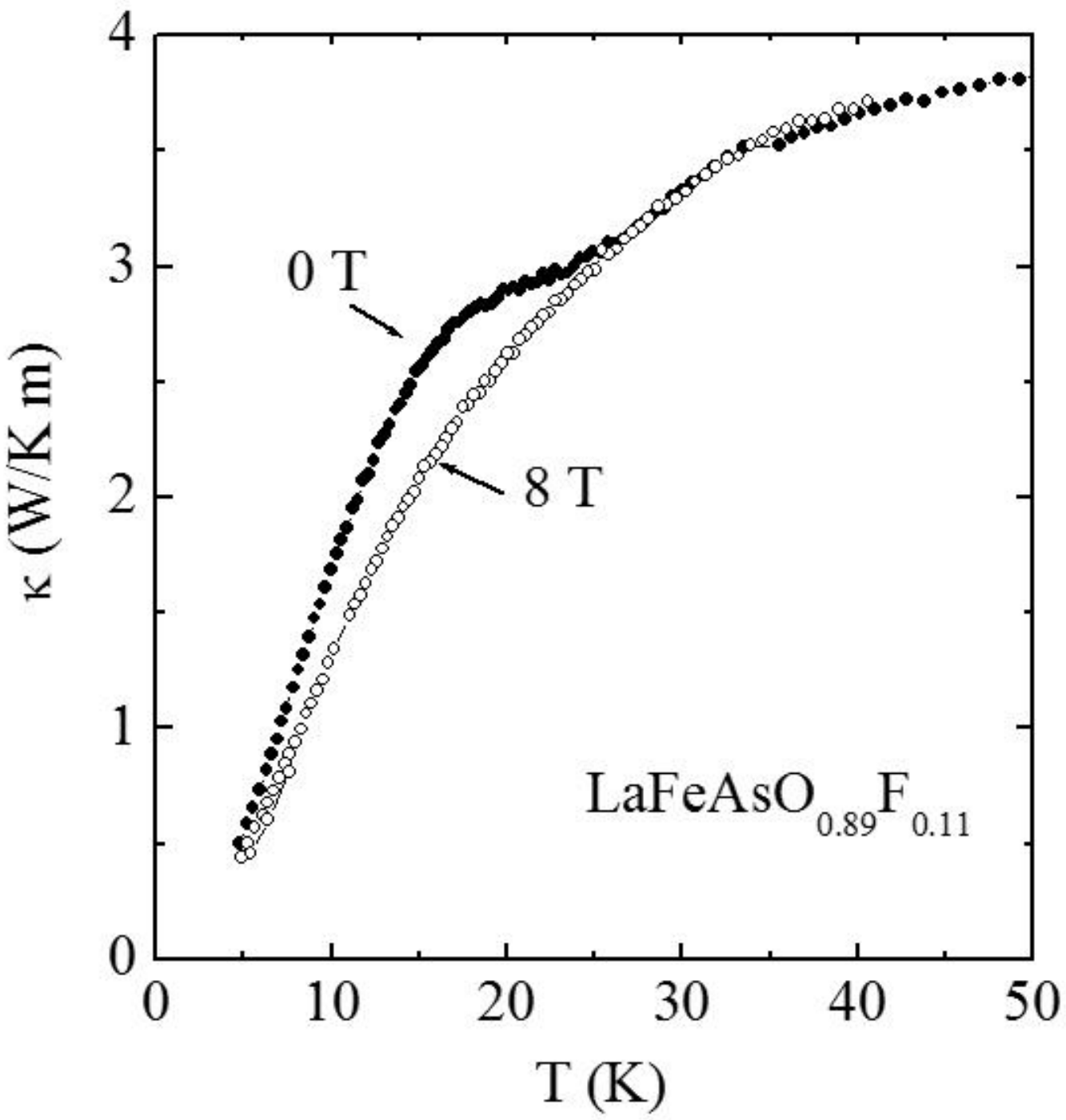

Figure 8: Temperature dependence of thermal conductivity data for $LaFeAsO_{0.89}F_{0.11}$ in applied fields of 0 T and 8 T.